\documentclass[12pt]{article}

\usepackage[pdfpagelabels]{hyperref}

\usepackage[usenames,dvipsnames]{color}
\usepackage{graphicx,epsfig,xcolor}

\hypersetup{colorlinks=true, linkcolor=violet, urlcolor=blue, citecolor=blue}

\usepackage{cite}
\usepackage[english]{babel}
\usepackage{amssymb,amsfonts,amsmath}
\usepackage{verbatim}
\usepackage{bbm,bbold,bm}
\usepackage{tensor}
\usepackage{slashed}
\usepackage{braket}

\def\q{\theta}
\def\a{\alpha}

\def\m{\mu}
\def\n{\nu}
\def\r{\rho}

\def\be{\begin{equation}}
\def\ee{\end{equation}}
\def\ba{\begin{eqnarray}}
\def\ea{\end{eqnarray}}
\def\nb{\nonumber}
\def\p{\partial}

\def\dc{\nabla}  
\def\a{\alpha}

\def\g{\gamma}
\def\G{\Gamma}
\def\d{\delta}

\def\l{\lambda}

\def\m{\mu}
\def\n{\nu}

\def\r{\rho}

\def\t{\tau}

\def\q{\quad}

\def\mc{\mathcal}

\newcommand{\pr}[1]{\left(#1\right)}

\usepackage{chngcntr}
\counterwithin*{equation}{section}

\usepackage{pifont}

\title{\bf Fracton gravity from spacetime dipole symmetry}

\author{\small Evangelos Afxonidis\footnote{afxonidisevangelos@uniovi.es},
Alessio Caddeo\footnote{caddeoalessio@uniovi.es}, 
Carlos Hoyos\footnote{hoyoscarlos@uniovi.es}, 
Daniele Musso\footnote{mussodaniele@uniovi.es} 
}

\date{}

\begin{document}

\maketitle
\vspace{-22pt}
\begin{center}\it{\small 
Department of Physics and Instituto de Ciencias y Tecnolog\'{\i}as Espaciales de Asturias (ICTEA) Universidad de Oviedo, c/ Leopoldo Calvo Sotelo 18, ES-33007 Oviedo, Spain
\\
}
\end{center}
\vspace{15pt}
\begin{abstract}
Dipole charge conservation forces isolated charges to be immobile fractons. These couple naturally to spatial two-index symmetric tensor gauge fields that resemble a spatial metric. We propose a spacetime Lorentz covariant version of dipole symmetry and study the theory of the associated gauge fields. In the presence of a suitable background field, these contain a massive anti-symmetric and a massless symmetric two-index tensors. The latter transforms only under longitudinal diffeomorphisms, making the massless sector similar to linearized gravity, but with additional modes of lower spin. We show that the theory can be consistently coupled to a curved background metric and study its possible interaction terms with itself and with matter. In addition, we construct a map between solutions of linearized gravity in Kerr-Schild form and solutions of fracton gravity coupled to matter.
\end{abstract}%

\newpage
\tableofcontents

\section{Introduction}
\label{sec:intro}

Theoretically possible new types of quantum phases of matter have been recognized in solvable lattice models such as the X-cube model and Haah's code \cite{Chamon:2004lew, Haah:2011drr,Yoshida:2013sqa,Vijay:2015mka,Vijay:2016phm,Williamson:2016jiq,Ma:2017aog,Halasz:2017xoq}. They display exotic properties such as excitations with restricted mobility, dubbed \emph{fractons}, and a vacuum degeneracy which is sensitive to the lattice size (see \emph{e.g.} \cite{Pretko:2020cko,Nandkishore:2018sel, Grosvenor:2021hkn, Gromov:2022cxa} for reviews). Among the possible realizations of fractonic dynamics, one of the simplest and most studied class is provided by systems conserving an Abelian charge and \emph{also} its dipole moment \cite{Pretko:2016kxt,Pretko:2016lgv,Seiberg:2019vrp}. An individual charge in isolation cannot move without changing the dipole moment, therefore the conservation of the latter forces the immobility of the former, leading to fractonic behavior.  

Dipole charge conservation may be implemented through a continuity equation involving a two-index symmetric current 
\begin{eqnarray}
    \partial_t \rho-\partial_i \partial_j J^{ij}=0\ ,
\end{eqnarray}
which in turn may be coupled to a symmetric tensor gauge field with two spatial indices $A_{ij}$. The tensor gauge field resembles the spatial components of a metric, suggesting thereby a connection between dipole conservation and spin-2 fields, and possibly gravity. For instance, in two-dimensional elasticity, disclinations are a type of lattice defect that can be interpreted as immobile conical singularities in the effective lattice geometry \cite{Kleinert:1989ky,Kleinert:2003za,Zaanen:2011hm,Beekman:2016szb,Zaanen:2021zqs,Tsaloukidis:2023bvz}. In the particle-vortex dual of such elasticity theory, the elastic degrees of freedom are described by a dynamical tensor gauge field and disclinations become (fractonic) charged particles  \cite{Pretko:2017kvd,Pretko_2018,Gromov:2019waa,Radzihovsky:2019jdo,Nguyen:2020yve,Gaa_2021,Tsaloukidis:2023bvz}. Thus fractons source the tensor gauge field producing a non-zero curvature in a similar way as matter sources the curvature of the metric in gravity. Another interesting connection among fractons and gravity was illustrated in \cite{Pretko:2017fbf}, where it was proposed that fracton (im)mobility could be understood as a realization of Mach's principle. Besides, many emergent theories of gravity \cite{Gu:2006vw,Gu:2009jh,Xu2006NovelAB,Xu:2010eg}  are fractonic, even when this was not recognized at first. Further studies of lattice models also suggest a deeper connection of fracton order and geometry\cite{Slagle:2017mzz}.

Although it could be maybe exaggerated to expect that a dipole-conserving theory captures all the features of gravity precisely, it might not be unreasonable to think of these type of theories as lying somewhere in between theories of ordinary matter, which are quite well understood, and gravity, which at the quantum level is still problematic. A systematic study of dipole or higher-multipole conserving theories could then give us a fresh perspective on gravity and higher-spin theories as well.

A clear difference between the tensor gauge theories mentioned so far and gravity is that the first are not Lorentz invariant, often they are not even formulated in a Lorentz covariant fashion. A possible way to bridge this difference is to promote the spatial tensor gauge field to a full tensor with two spacetime indices $A_{\mu\nu}$ and try to formulate the theory covariantly. This was the path followed recently in \cite{Bertolini:2022ijb, Blasi:2022mbl,Bertolini:2023juh}, where a quadratic action was constructed including the linearized Einstein tensor and additional terms allowed by the smaller gauge symmetry of the tensor field $\delta A_{\mu\nu}=\partial_\mu\partial_\nu \alpha$ as compared to the linearized diffeomorphism symmetry of the metric $\delta h_{\mu\nu}=\partial_{(\mu}\xi_{\nu)}$.

Despite the direct approach being fine in flat spacetime, one might be interested in coupling the theory to a non-trivial background metric, and in this case the two-derivative gauge transformation of the tensor field would clash with covariance under background diffeomorphisms, an obstruction already observed in the non-Lorentz covariant context \cite{Gromov:2017vir,Slagle:2018kqf,Jain:2021ibh, Bidussi:2021nmp}. Another puzzling aspect of the tensor gauge theory is that the gauge fixing seems to require a single scalar gauge condition but a vector gauge fixing is also possible, and according to \cite{Bertolini:2023juh} even preferred. We will address these issues by generalizing the approach described in 
 \cite{Caddeo:2022ibe}. There we proposed an improvement in the formulation of dual elasticity by realizing dipole and higher-moment transformations through internal symmetries. This approach allowed us to replace tensor gauge fields by ordinary vector fields, which circumvents the issues in the coupling to a background geometry. The tensor gauge theory is recovered from the vector field formulation by a partial gauge fixing, yielding equivalent results for elasticity in flat space.

The present paper is structured as follows. In Section \ref{sec:symmandfields}, we generalize the monopole dipole moment algebra (MDMA) to the Lorentz covariant case and introduce the corresponding gauge fields and associated curvatures. In Section \ref{sec:quadaction}, we present the most general quadratic action both in the fields and in the derivatives for the spacetime dipole gauge theory. We show that the theory of the covariant tensor gauge field of \cite{Blasi:2022mbl,Bertolini:2023juh} is recovered as a special case. We re-examine the questions of the number of degrees of freedom and gauge fixing, as well as the connection with linearized gravity, and show how the theory can be naturally coupled to a background metric. In Section \ref{secinteractions}, after providing details on dipole conservation in the covariant case, we explore the possible interactions allowed by gauge invariance --self-couplings of the tensor field and couplings to matter-- and find a map between solutions of linearized gravity and solutions of the tensor gauge theory coupled to matter. We conclude with an outlook and with two appendices containing  technical details of the computations presented in the main text.

\section{Symmetries and gauge fields}
\label{sec:symmandfields}

Let us describe the Lorentz covariant generalization of the gauge theory proposed in \cite{Caddeo:2022ibe} (see also \cite{Radzihovsky:2019jdo} for a similar realization). We use Greek letters for spacetime indices and capital Latin ones for internal indices. Most of our considerations will be valid when both the spacetime and the internal space have arbitrary dimension $D=d+1$, although we will focus on $D=4$ when considering explicit solutions as in Subsections \ref{subsecsolutionsymmetricfield} and \ref{subsecmaptolinearized}. Consider the generators of internal translations $P_A$, Abelian $U(1)$ symmetry $Q$, and a vector generator $Q^A$, with non-trivial commutator
\begin{equation}
i[P_A,Q^B]=\delta_A^B Q \ .
\end{equation}
When $A$ and $B$ are constrained to internal spatial indices this constitutes the usual monopole-dipole-momentum algebra discussed in \cite{Gromov:2018nbv,Pena-Benitez:2021ipo} in the context of dipole charge conservation and fractons. The algebra above is an extension of the spatial MDMA including internal time components of the vector currents.%

Following \cite{Caddeo:2022ibe}, we introduce a gauge connection for the extended MDMA
\begin{equation}
{\cal A}_\mu= e_\mu^{\ A}P_A+a_\mu Q+ b_{\mu A} Q^A\ .
\end{equation}
An infinitesimal gauge transformation has the usual form
\begin{equation}
\delta_\Lambda {\cal A}_\mu =D_\mu \Lambda=\partial_\mu \Lambda+i[{\cal A}_\mu,\Lambda]\ ,
\end{equation}
where the gauge parameter, expanded in the algebra generators, is parameterized as
\begin{equation}
\Lambda=\xi^A P_A+\lambda_0 Q+\lambda_{1A} Q^A \ .
\end{equation}
The components of the connection thus transform as
\begin{subequations}
\label{fieldsgaugetransform}
\ba 
\delta e_{\mu}^{\ A} &=& \partial_\mu\xi^A \  ,\\
\delta a_\mu &=& \partial_\mu\lambda_0+e_\mu^{\ A}\lambda_{1A}-b_{\mu A}\xi^A \ ,\\
\delta b_{\mu A}&=& \partial_\mu\lambda_{1\,A} \ .
\ea
\end{subequations}
Defining the field strength of $a_\mu$ as $f_{\mu\nu}=\partial_\mu a_\nu-\partial_\nu a_\mu$, there are two curvature invariants
\begin{subequations}
\label{defHmunuTmunu}
\ba
T_{\mu\nu}^A &=&\p_\mu e_\nu^{\ A}-\p_\nu e_\mu^{\ A} \ ,\\
H_{\mu\nu A} &=& \p_\mu b_{\nu A}-\p_\nu b_{\mu A} \ .
\ea
\end{subequations}
and a covariant curvature
\begin{equation}
\label{defBmunu}
    B_{\mu\nu} = b_{\mu A}e_\nu^{\ A}-b_{\nu A} e_\mu^{\ A}-f_{\mu\nu} \ ,
\end{equation}
transforming as
\begin{equation}
    \delta B_{\mu\nu}
    =
    - T_{\mu\nu}^A \lambda_{1\, A}
    + H_{\mu\nu\, A}\xi^A\ .
\end{equation}
Introducing the spacetime and internal metrics $\eta_{\mu\nu}$ and $\eta_{AB}$, it is possible in principle to construct a fully gauge-invariant and Lorentz-covariant theory for these fields.

In the following, we treat the gauge field associated to the internal translations $e_\mu^{\ A}$ as a background field 
and consider only $a_\mu$ and $b_{\mu A}$ as dynamical fields. In order to have non-trivial gauge transformations and preserve Lorentz covariance, we fix%
\footnote{In principle all the gauge fields have mass dimension one, whereas the gauge transformation parameters are dimensionless. When  selecting the background there is an overall physical scale $M$ entering through $e_\mu^{\ A} = M \delta_\mu^{\ A}$ that we have fixed to one.}
\begin{eqnarray}\label{cho:emu}
e_\mu^{\ A}=\delta_\mu^{\ A}\ .    
\end{eqnarray}
The background field fixed in this way breaks the product of spacetime and internal Lorentz transformations to the diagonal subgroup. 
The background field is akin to a vielbein that we can use to convert all the internal indices to spacetime indices, although this in general breaks the internal transformations generated by $\xi^A$, unless one considers Lagrangians constructed out of the invariant combinations such as \eqref{defHmunuTmunu} only.

In the following, we allow the breaking of the internal translations generated by $\xi^A$.
Upon taking the choice \eqref{cho:emu}, the dynamical fields transform as
\begin{subequations}
\label{eq:gaugetransf}
\ba 
\delta a_\mu &=& \partial_\mu\lambda_0+\lambda_{1\mu} \ ,\\
\delta b_{\mu A}&=& \partial_\mu\lambda_{1\,A} \ .
\ea
\end{subequations}
Under this restricted symmetry, with $T_{\mu\nu}^A=0$, $B_{\mu\nu}$ becomes an additional curvature invariant. In addition, we can construct a new curvature invariant with one derivative acting on the $b_{\mu A}$ field%
\footnote{For $f_{\mu\nu}=0$ and $b_{\mu\nu}=b_{\nu\mu}$, the tensor $\Gamma_{\mu\nu\rho}$ reduces to the field strength introduced in \cite{Bertolini:2022ijb,Bertolini:2023sqa}.}
\ba
\Gamma_{\mu\nu\lambda}  &=& \partial_ \mu (b_{\nu\lambda}-f_{\nu\lambda})+\partial_\nu (b_{\mu\lambda}-f_{\mu\lambda})-\partial_\lambda (b_{\mu\nu}+b_{\nu\mu}) \nb \\
&=&
2\left[\partial_{(\mu}B_{\nu)\lambda}
- \frac{1}{2}\left( H_{\lambda \mu A} e_{\nu}^{\ A}+H_{\lambda \nu A} e_{\mu}^{\ A}\right)
\right]
\,.
\ea
Since the last index of $H_{\mu\nu A}$ is contracted with $e_\mu^{\ A}$, the curvature $\Gamma_{\mu\nu\lambda}$ is not invariant under internal transformations generated by $\xi^A$. From now on, excepting Subection \ref{seccurvedmetric}, we will allow such symmetry to be explicitly broken by the background \eqref{cho:emu} and accordingly we use only spacetime indices in all quantities.

Finally, it is also interesting to note that all the curvature invariants depend on the same combination of the gauge fields $b_{\mu\nu}-\p_\mu a_\nu$
\begin{subequations}
\label{eq:invcurvatures}
    \ba
H_{\mu\nu \lambda} &=& \p_\mu( b_{\nu\lambda}-\p_\nu a_\lambda)-\p_\nu (b_{\mu\lambda}-\p_\mu a_\lambda)\ ,\\
B_{\mu\nu}&=&b_{\mu\nu}-\p_\mu a_\nu-(b_{\nu\mu}-\p_\nu a_\mu)\ ,\\
\Gamma_{\mu\nu\lambda}&=&\p_\mu(b_{\nu\lambda}-\p_\nu a_\lambda)+\p_\nu(b_{\mu\lambda}-\p_\mu a_\lambda)-\p_\lambda(b_{\mu\nu}-\p_\mu a_\nu+b_{\nu\mu}-\p_\nu a_\mu)\ .
    \ea
\end{subequations}
This fact will be of relevance for the derivation of the equations of motion and discussion thereof.

\section{Quadratic action and massless solutions}
\label{sec:quadaction}

The most general action exhibiting invariance under spacetime Poincaré, parity and time-reversal symmetry, and being quadratic with at most two derivatives acting on $b_{\mu A}$ is\footnote{One could also consider the terms $H_{\m \n \r} H^{\m \r \n}$ and $\G_{\m \n \r} \G^{\m \r \n}$. However, it can be shown using integration by parts that they can be written as linear combinations of the terms considered in (\ref{lag_quad_2}). We thank the anonymous referee for pointing out this and the possibility of adding $\alpha_6$ and $\alpha_7$ terms.}
\ba
\label{lag_quad_2}
{\cal L}&=&-\frac{\alpha_1}{4} H_{\mu\nu\lambda}H^{\mu\nu\lambda}-\frac{\alpha_2}{4} B_{\mu\nu} B^{\mu\nu}-\frac{\alpha_3}{4}\Gamma_{\mu\nu\lambda}\Gamma^{\mu\nu\lambda}-\frac{\alpha_4}{4}\Gamma_\mu^{\ \mu \lambda}\Gamma^\nu_{\ \nu \lambda} 
- \frac{\alpha_5}{4} H_{\mu\nu\lambda} \Gamma^{\mu\lambda\nu}
\nonumber \\ 
 &-&\frac{\alpha_6}{4} \Gamma_{\lambda\mu}^{\ \ \mu}H^{\lambda\nu}_{\ \ \nu}
-\frac{\alpha_7}{4}H_{\lambda\mu}^{\ \ \mu}H^{\lambda\nu}_{\ \ \nu}+b_{\mu\nu}J^{\mu\nu}+a_\mu J^\mu
\ ,
\ea
with $\alpha_1,\dots,\alpha_7$ coefficients weighting the quadratic curvature terms. We also included the coupling to the sources $J^{\mu\nu}$ and $J^\mu$. 

In order for the action to be gauge-invariant, the sources must satisfy the (non-)conservation equations
\begin{equation}
\label{eq:ward}
\p_\mu J^\mu =0 \ , \qquad \p_\mu J^{\mu\nu}=J^\nu\ .   
\end{equation}
The combination of the two implies that
\begin{equation}\label{eq:wardf}
    \p_\mu \p_\nu J^{\mu\nu}=0\ .
\end{equation}
The second equation in \eqref{eq:ward} allows us to rewrite the coupling to the currents in the actions as a single coupling to the tensor current
\begin{equation}
    b_{\mu\nu} J^{\mu\nu}+a_\mu J^\mu\longrightarrow \left(b_{\mu\nu}-\p_\mu a_\nu \right)J^{\mu\nu}\ .
\end{equation}
Together with \eqref{eq:invcurvatures}, this shows that the action only depends on the combination $b_{\mu\nu}-\p_\mu a_\nu$. As a consequence, the equations of motion for $a_\mu$ are not independent, but they are equal to the divergence of the equations of motion for $b_{\mu\nu}$. Thanks to this fact we can perform a field redefinition
\begin{equation}
\label{fieldredefinition}
    b_{\mu\nu}=\partial_\mu a_\nu+\frac{1}{2}\left(B_{\mu\nu}+h_{\mu\nu}\right) \ ,
\end{equation}
where $B_{\mu\nu}$ is an antisymmetric tensor (which coincides with the curvature $B$) and $h_{\mu\nu}$ is symmetric. The equations of motion for $B$ and $h$ are respectively equal to the antisymmetric and symmetric parts of the equations of motion for $b_{\mu\nu}$. The full set of equations of motion for each field can be found in Appendix~\ref{app:eoms}.

Consistency with the gauge transformations \eqref{eq:gaugetransf} demands that the symmetric field transforms in the following way:
\begin{equation}\label{eq:gaugetrh}
    \delta h_{\mu\nu}=-2\p_{\mu}\p_{\nu} \lambda_0\ .
\end{equation}
This coincides with the transformation of the fracton gauge field proposed by \cite{Blasi:2022mbl}, and determines largely the structure of the action for the symmetric field, as they discuss in detail.

In our case, the action9 can be split in three pieces
\begin{equation}
    {\cal L}= {\cal L}_{B^2}+{\cal L}_{hB}+{\cal L}_{h^2}\ .
\end{equation}
The first piece is the action for the antisymmetric field 
\ba
    \label{onl_bee}
    {\cal L}_{B^2} 
    &=&
    - z_{1} \p_\rho B_{\mu\nu}\p^\rho B^{\mu\nu} \nb 
    -  z_{2} \p^\nu B_{\nu\rho} \partial_\mu B^{\mu\rho}
    -\frac{\alpha_2}{4}B_{\mu\nu}B^{\mu\nu} 
    \ ,
\ea
where
\be
z_{1} = \frac{1}{16}\left(
    2\alpha_1 + 2\alpha_3 - \a_{5}\right) \ , \q \q z_{2} = \frac{1}{16}\left(-2\alpha_1
    + 2\alpha_3
    + 4\alpha_4
    - \alpha_5
    -\alpha_6+\alpha_7
    \right) \ .
\ee
This is similar to a massive two-form. Next, there is a piece coupling the symmetric and antisymmetric fields
\begin{equation}
    {\cal L}_{hB}
    = 
     z_{3} \p_\mu h_{\nu\rho} \partial^\nu  B^{\mu\rho} + z_{4} \p_{\m} h \p_{\n} B^{\m \n} \ ,
\end{equation}
where
\be
z_{3} = \frac{1}{4}\left(\a_{1} -  3\alpha_3 - 2\alpha_4 
    +\frac{1}{2}\alpha_7 \right) \ ,\q \q \q z_{4} = - \frac{4 \a_{4} - \a_{7}}{8} \ .
\ee

The remaining piece is the action for the symmetric field. It will be convenient to introduce the gauge-invariant tensors
\begin{equation}\label{eq:HG}
\begin{split}
& H^\mu=\partial_\sigma h^{\sigma \mu}-\partial^\mu h\ ,\\
& G^{\mu\nu}=\partial^2 h^{\mu\nu}-\partial^\mu \partial^\nu h -(\partial^\mu H^\nu+\partial^\nu H^\mu)+\eta^{\mu\nu}\partial_\sigma H^\sigma\ .   \end{split}
\end{equation}
The tensor $G$ satisfies the relations
\begin{equation}
    \partial_\mu G^{\mu\nu}=0 \ ,\qquad \eta_{\mu\nu}G^{\mu\nu}=(d-2)\partial_\sigma H^\sigma\ ,
\end{equation}
and it is the analog to the Einstein tensor in linearized gravity. In fact, it is invariant under the larger set of transformations corresponding to linearized diffeomorphisms $\delta h_{\mu\nu}=\partial_\mu \xi_\nu+\partial_\nu \xi_\mu$.

Expressed in terms of these tensors, the action takes the simple form
\begin{equation}\label{lag_h2}
    {\cal L}_{h^2} = (g_1-g_2) h_{\mu\nu} G^{\mu\nu}+ g_2 H_\mu H^\mu  \ ,
\end{equation}
where we have used the conventions introduced in \cite{Bertolini:2023juh} to define the couplings
\begin{align}
    \frac{4 \alpha_4 + \a_{6} + \a_{7}}{16} &= -g_1\ ,\\ \label{beta}
    \frac{1}{16}\left(2\alpha_1+6\alpha_3+3\alpha_5\right)
    \equiv
    \frac{\beta}{8}
    &= g_1-g_2\ .
\end{align}
Up to this point, the theory is defined by seven parameters, that after the redefinition (\ref{fieldredefinition}) we rewrote as $\{g_{1},g_{2},\a_{2},z_{1},z_{2},z_{3},z_{4}\}$. However, the parameter $z_{4}$ is redundant, as it multiplies a Lagrangian term that vanishes identically after integration by parts, so the theory is in fact characterized by six independent parameters. A consistent choice would be to just set $\alpha_7=0$ for instance.\\
In the following, it will be useful to fix one parameter to decouple $h_{\m \n}$ from $B_{\m \n}$, imposing $z_{3}=0$, namely
\be
 \a_{1} -  3\alpha_3 - 2\alpha_4 
    +\frac{1}{2}\alpha_7  = 0 \ .
\ee
Moreover, we can fix another parameter to obtain that the kinetic term of the antisymmetric field become that of a massive two-form. Let us define
\begin{equation}\label{cal_H}
    {\cal H}_{\mu\nu\lambda}=3\partial_{[\mu} B_{\nu \lambda]}=\partial_\mu B_{\nu\lambda}+\partial_\nu B_{\lambda\mu}+\partial_\lambda B_{\mu\nu}\ .
\end{equation}
Imposing 
\be
z_{1} = - \frac{z_{2}}{2} \equiv \frac{g_{3}}{2} \ ,
\ee
the Lagrangian $\mc{L}_{B^{2}}$ reads
\begin{equation}\label{eq:Baction}
    {\cal L}_{B^2}= - \frac{g_3 }{6}{\cal H}_{\mu\nu\lambda}{\cal H}^{\mu\nu\lambda}-\frac{\alpha_2}{4}B_{\mu\nu} B^{\mu\nu}\ .
\end{equation}
As a result, after imposing these two conditions, the theory is defined by four independent parameters: $g_{1}$, $g_{2}$, $g_{3}$, $\a_{2}$.

After decoupling the antisymmetric field, the equations of motion for the symmetric field are
\begin{equation}\label{eq:eqh}
2(g_1-g_2)\, G^{\mu\nu}
   -g_2\left(
   \partial^\mu H^\nu
   +\partial^\nu H^\mu
   -2 \eta^{\mu\nu}\partial_\sigma H^\sigma \right) +\frac{1}{2}J^{(\mu\nu)}=0\ ,
\end{equation}
where we used the standard fact that $h_{\mu\nu}G^{\mu\nu}$ maps to $G_{\mu\nu}h^{\mu\nu}$ upon double integration by parts.

\subsection{Solutions for the symmetric field}
\label{subsecsolutionsymmetricfield}

Let us determine what the massless degrees of freedom of our theory are.
From now on, we will work in $D=4$ dimensions and we consider the case in which the symmetric field is decoupled from the antisymmetric field, corresponding to the condition
\begin{equation}\label{eq:decoupling}
    \alpha_1=3\alpha_3+2\alpha_4 -\frac{1}{2}\alpha_7\ .
\end{equation}
The solutions for the decoupled massive anti-symmetric field are straightforward to obtain. Here we focus on the more subtle analysis of the symmetric field $h_{\mu \nu}$ solutions to equation (\ref{eq:eqh}).
First we expand in plane waves of momentum $k^\mu$
\begin{equation}
    h_{\mu\nu}(x)=\int \frac{d^4 k}{(2\pi)^4}\, \varepsilon_{\mu\nu}(k) e^{ik\cdot x}\ ,
\end{equation}
where $\varepsilon_{\mu\nu}$ is a polarization tensor. The invariance of the action under the gauge transformations \eqref{eq:gaugetransf} implies that the pure-gauge polarization $\varepsilon_{\mu\nu}=k_\mu k_\nu$ is a solution for any momentum. The polarization tensor for solutions that are not pure gauge can then be constrained to be transverse
\begin{equation}\label{eq:tranvcond}
    k^\mu k^\nu \varepsilon_{\mu\nu}(k)=0\ .
\end{equation}
The equations of motion admit non-trivial solutions for massless momenta. We can construct the polarization tensors using a basis of polarization vectors $e^\sigma_\mu$, $\sigma=\pm 1$ ($e^{-\sigma}=\overline{e}^{\ \sigma}$), $q_\mu$ satisfying (the bar denotes complex conjugation)
\begin{equation}
\label{orto}
e^\sigma\cdot e^{\sigma}=k\cdot e^{\sigma}=q\cdot e^\sigma=0 \ , \quad \quad ik\cdot q=1 \ , \quad \quad \overline{e}^{\,\sigma}\cdot e^{\sigma'}=\delta^{\sigma \sigma'}\ . .
\end{equation}
For instance, if we take the direction of the spatial momentum as the $z$-axis,
\begin{equation}
\label{ref_bas}
\begin{split}
&k_\mu=\left(\begin{array}{cccc} -\omega & 0 & 0 &  k_z  \end{array} \right) \ , \\
&q_\mu=-\frac{i}{\omega^2+k_z^2}\left(\begin{array}{cccc} \omega & 0 & 0 &  k_z \end{array} \right) \ , \\\ 
&e^{\pm 1}_\mu=\frac{1}{\sqrt{2}}\left(\begin{array}{cccc} 0 & 1 & \mp i &  0 \end{array} \right) \ .
\end{split}
\end{equation}
Any other choice of momentum and polarization vectors can be obtained by applying spatial rotations to the vectors above. Restricting to positive frequency solutions, the on-shell momentum corresponds to $\omega=|k_z|$, in which case $k^2=q^2=0$. 

We decompose the Fourier transform of $h_{\mu\nu}$ with null momentum $k^\mu$ as follows
\begin{equation}\label{mod_exp}
    \varepsilon_{\mu\nu}=\varepsilon_{\mu\nu}^{t\perp}+\varepsilon_{\mu\nu}^\perp+\varepsilon_{\mu\nu}^t \ ,
\end{equation}
where ``${}^t$'' means traceless and ``${}^\perp$'' means transverse. In terms of our basis
\begin{equation}
\label{spli_acc}
\begin{split}
&    \varepsilon_{\mu\nu}^{t\perp}=\sum_\sigma A_\sigma e^\sigma_\mu e^\sigma_\nu+\sum_\sigma A_{3\sigma} \frac{i k_{(\mu}e^\sigma_{\nu)}}{\omega}+A_4 \frac{k_\mu k_\nu}{\omega^2} \ ,\\
&  \varepsilon_{\mu\nu}^\perp=B\left( \eta_{\mu\nu}-2i k_{(\mu}q_{\nu)}\right) \ ,\\
&\varepsilon_{\mu\nu}^t=C_1\omega^2 q_\mu q_\nu+\sum_\sigma C_{2\sigma}\omega q_{(\mu}e^\sigma_{\nu)}+C_3\left( \eta_{\mu\nu}-4 ik_{(\mu}q_{\nu)}\right)\ .
\end{split}
\end{equation}
The $A_4$ term automatically cancels in the equations of motion for any momenta, thus it corresponds to a pure gauge component. Otherwise non-trivial solutions only exist for null momenta $k^2=0$.

A straightforward calculation (see Appendix~\ref{app:eoms}) for the case without mixing with the antisymmetric field fixes $C_1=C_{2\sigma}=0$ and 
\begin{equation}
    B=-\frac{2 g_1-g_2}{2g_1}\, C_3\ .
\end{equation}
Then, the independent modes have the polarizations
\begin{equation}\label{table}
    \begin{array}{c|c|c}
\text{helicity}\ A & \varepsilon_{\mu\nu}^A &\text{pure\; gauge}\\ \hline
 2\sigma & e_ \mu^\sigma e_\nu^\sigma &g_1=g_2\\
 \sigma & ik_{(\mu}e_{\nu)}^\sigma & g_2=0\\
 0 &  g_2\eta_{\mu\nu}-2(2 g_1+g_2) i k_{(\mu}q_{\nu)} & g_2=0,\ g_2=-2g_1
    \end{array}
\end{equation}
Note that $\eta^{\mu\nu}\varepsilon_{\mu\nu}^0=-2(2g_1-g_2)$ vanishes for $g_2=2g_1$, and that for this choice the kinetic term of the antisymmetric field in \eqref{eq:Baction} vanishes as well, so $B_{\mu\nu}$ becomes non-dynamical.

The helicity zero tensor does not satisfy the transversality condition \eqref{eq:tranvcond} off-shell, in principle one could add a term proportional to $k^\mu k^\nu/k^2$ to ensure this condition, but it does not have a good massless limit $k^2\to 0$.

If the condition \eqref{eq:decoupling} is not satisfied, one has to take into account the coupling to the antisymmetric field. However, one can check that this does not change the massless solutions for generic values of the couplings $g_1$ and $g_2$. If the antisymmetric field is decoupled, the solutions do change for some special values of the couplings $g_1$ and $g_2$, as also discussed in \cite{Bertolini:2023juh} from a different perspective. For $g_2=0$ the action of the symmetric field becomes that of linearized gravity and there is a larger diffeomorphism invariance. In this case, the modes of helicity $A=\pm 1$ and $A=0$ become pure gauge%
\footnote{By this we mean that they automatically satisfy the equations of motion without imposing any condition on the momenta, as it was also the case for the $A_4$ term in \eqref{table}.}.
For $g_2=-2 g_1$ the action becomes independent of the trace, and the helicity $A=0$ mode becomes pure gauge. Finally, for $g_1=g_2$, the part of the action proportional to that of linearized gravity is removed and the helicity $A=\pm 2$ modes become pure gauge, see \eqref{lag_h2}.

In the following we will only study the theory for values of the couplings that do not alter the degrees of freedom of the theory, \emph{i.e.} $g_2\neq \{ 0,g_1,\pm 2 g_1\}$.

\subsection{Gauge-fixing and propagator}
\label{pro:sec}

In order to obtain a propagator for the symmetric field $h_{\mu \nu}$, one needs to introduce a gauge fixing that removes unphysical degrees of freedom. In principle, since the transformation \eqref{eq:gaugetransf} involves a single function, there should be a single (scalar) gauge fixing condition, however the authors of  \cite{Bertolini:2023juh} found that an alternative vector gauge fixing condition also seems to work consistently.

Our realization in terms of the spacetime dipole symmetry is useful to understand this matter. Following the procedure of BRST quantization, we can introduce a set of ghosts, anti-ghosts and auxiliary fields for the monopole ($c$, $\bar{c}$, $b$) and dipole ($c_A$, $\bar{c}^A$, $b^A$) transformations. The nilpotent BRST transformations $\mathfrak{s}$ are 
\begin{equation}
    \begin{array}{lclclcl}
\mathfrak{s}\, a_\mu &= &\partial_\mu c+\delta_\mu^A c_A \ , & \ & \mathfrak{s}\, b_{\mu A} &=& \partial_\mu c_A \ ,\\
\mathfrak{s}\, c &=& 0 \ , & \ & \mathfrak{s}\, c_A &=& 0 \ ,\\
\mathfrak{s}\, \bar{c} &=& b \ , & \ & \mathfrak{s}\, \bar{c}^A &=& b^A \ ,\\
\mathfrak{s}\, b &= &0 \ , & \ & 
\mathfrak{s}\, b^A &=& 0 \ .
    \end{array}
\end{equation}
As gauge fixing we introduce a BRST-exact term in the action
\begin{equation}
    {\cal L}_{g.f.}=-\mathfrak{s}\, W \ ,
\end{equation}
where among the possible choices there could be scalar gauge fixings, such as
\begin{equation}\label{eq:scalargf}
    W_s=\bar{c}\,\left[\partial^\mu\partial^\nu(b_{\mu A}\delta_\nu^A-\partial_\mu a_\nu)-\frac{\xi}{2} b\right]+\bar{c}^A \left[a_\mu\delta_A^\mu-\frac{\kappa}{2} \eta_{AB} b^B\right] \ ,
\end{equation}
or vector gauge fixings like
\begin{equation}\label{vec:gf}
    W_v=\bar{c}\,\left[\partial^\mu a_\mu-\frac{\xi}{2}b\right]+\bar{c}^A\left[\partial^\mu b_{\mu A}-\frac{\kappa}{2}\eta_{AB} b^B\right] \ .
\end{equation}
The most convenient choice in our case is \eqref{eq:scalargf}, since it will fix $a_\mu$ and remove the pure gauge modes of the symmetric field without affecting  other modes. The gauge fixing action is
\begin{equation}
    {\cal L}_{g.f.}=-b\partial^\mu\partial^\nu(b_{\mu \nu}-\partial_\mu a_\nu)+\frac{\xi}{2}b^2-b^\mu a_\mu+\frac{\kappa}{2} b_\mu b^\mu+\text{ghosts}\ .
\end{equation}
Integrating out the auxiliary fields leads to the terms
\begin{equation}\label{eq:gaugefixscalar}
    {\cal L}_{g.f.}'=-\frac{1}{2\xi}\left[\partial^\mu\partial^\nu(b_{\mu \nu}-\partial_\mu a_\nu)\right]^2-\frac{1}{2\kappa}a_\mu a^\mu+\text{ghosts}\ .
\end{equation}
The equation of motion for $a_\mu$ equals the divergence of the equation of motion for $b_{\mu\nu}$, except for the contribution originating in the gauge fixing term depending on $\kappa$, the only place where $a_\mu$ does not enter the Lagrangian through the combination $b_{\mu\nu} - \partial_\mu a_\nu$. Such term forces the condition $a_\mu=0$ for any $\kappa$. On the other hand, the gauge-fixing term depending on $\xi$ removes the pure-gauge polarization of the symmetric field, namely the one whose polarization tensor $\varepsilon_{\mu\nu}$ is $k_\mu k_\nu$, except for $k^2=0$. The other modes of the field $b_{\mu\nu}$ are not affected by its presence. 

The Fourier transform of the equations of motion for the symmetric field is
\begin{equation}
\Delta^{\mu\nu,\alpha\beta}\varepsilon_{\alpha\beta}=\frac{1}{2}\tilde{J}^{\mu\nu}\ ,
\end{equation}
where $\tilde{J}^{\mu\nu}$ is the Fourier transform of the current. The tensor $\Delta$ is
\begin{equation}\label{eq:eqh:fou}
    \Delta^{\mu\nu,\alpha\beta}=(g_1-g_2)k^2 I^{\mu\nu,\alpha\beta}+\frac{1}{2}(g_2-2g_1)K_1^{\mu\nu,\alpha\beta}-2 g_1 K_2^{\mu\nu,\alpha\beta}+\frac{1}{4\xi}k^\mu k^\nu k^\alpha k^\beta\ .
\end{equation}
Here, we have defined the tensors
\begin{equation}\label{ten:def}
\begin{split}
    & I^{\mu\nu,\alpha\beta}=\eta^{\mu\alpha}\eta^{\nu\beta}+\eta^{\mu\beta}\eta^{\nu\alpha} \ ,\\
    & K_1^{\mu\nu,\alpha\beta}=k^\mu k^\alpha \eta^{\nu\beta}+k^\mu k^\beta \eta^{\nu\alpha}+k^\nu k^\alpha \eta^{\mu\beta}+k^\nu k^\beta \eta^{\mu\alpha} \ ,\\
    & K_2^{\mu\nu,\alpha\beta}=k^2\eta^{\mu\nu}\eta^{\alpha\beta}-k^\alpha k^\beta\eta^{\mu\nu}-k^\mu k^\nu \eta^{\alpha\beta}\ .
    \end{split}
\end{equation}
The propagator $G$ is the inverse of $\Delta$ over symmetric tensors
\begin{equation}\label{green}
    \Delta^{\mu\nu,\alpha\beta}G_{\alpha\beta,\sigma\rho}=\frac{1}{4}\left( \delta^\mu_\sigma \delta^\nu_\rho+\delta^\mu_\rho\delta^\nu_\sigma\right)\ .
\end{equation}
We find
\begin{equation}\label{prop}
\begin{split}
    G_{\alpha\beta,\sigma\rho}=&\frac{1}{8(g_1-g_2)}\frac{1}{k^2}\left[ I_{\alpha\beta,\sigma\rho}+\frac{g_2-2g_1}{g_2}\frac{1}{k^2}K_{1\,\alpha\beta,\sigma\rho}-\frac{2g_1}{2g_1+g_2}\frac{1}{k^2}K_{2\, \alpha\beta,\sigma\rho}\right]\\
    &+\frac{1}{k^2}\left(\frac{1}{g_2}+\frac{g_1+g_2}{4(g_1-g_2)(2g_1+g_2)}+\frac{2\xi}{k^2} \right) \frac{k_\alpha k_\beta k_\sigma k_\rho}{(k^2)^2} \ .
    \end{split}
\end{equation}
Let us point out that, for any of the special values of the coupling where some modes become pure gauge \eqref{table}, the propagator has a divergent coefficient, so there is no smooth limit from the general theory to those special cases. Since the gauge parameter $\xi$ is present only in the completely longitudinal term of the propagator, the projection over the polarizations of the physical modes is independent of $\xi$. More generally, the unphysical double pole in the propagator with coefficient $\xi$, which might be worrisome in other theories with higher derivative terms in the action, would not contribute to any gauge-invariant observable.

\subsection{Coupling to curved geometry}
\label{seccurvedmetric}

As mentioned in the introduction, the rank-two tensor realization of fracton gauge theories exhibits some issues when attempting to couple these theories to a curved background geometry. Indeed, in that case, the dipole gauge transformation involves the second derivative of a parameter and, as such, does not admit a natural covariant counterpart. 
In this section, we elaborate on how our realization permits such a coupling. Since the present paper focuses on the relativistic case, we describe the coupling to a pseudo-Riemannian geometry, but extending it to non-relativistic contexts is in principle straightforward.

Let us call $\g_{\m \n}$ the metric of the background geometry, which is independent of our dynamical gauge field $h_{\m \n}$, and
$\dc_\m$ the covariant derivative defined with the Christoffel symbols computed from $\g_{\m \n}$. With these ingredients, we can make the theory invariant under background diffeomorphisms
\be
\d \g_{\m \n} = \dc_{\m} \zeta_{\n} + \dc_{\n} \zeta_{\m} \ .
\ee
When doing this, we need to remember that some of the spacetime indices originate from internal indices after choosing the background $e_\m ^{\ A}$, hence it is better to work with the formul$\ae$ that still involve internal indices explicitly. The gauge fields of our theory are one-form connections, and, therefore, transform under background diffeomorphisms as the Lie derivative,
\begin{subequations}
\ba
\delta_\zeta e_\mu ^{\ A} &= {\cal L}_\zeta e_\mu ^{\ A} &=\zeta^\r \partial_\r e_\mu ^{\ A} + e_\r ^{\ A} \partial_\mu\zeta^\r  \ ,\\
\delta_\zeta a_\mu &= {\cal L}_\zeta a_\mu&=\zeta^\r \partial_\r a_\mu+ a_\r \partial_\mu\zeta^\r  \ ,\\
\delta_\zeta b_{\mu A} &= {\cal L}_\zeta b_{\mu A}&=\zeta^\r \partial_\r b_{\mu A}+ b_{\r A} \partial_\mu\zeta^\r
\ .
\ea
\end{subequations}

Since internal indices are not involved in diffeomorphism transformations, the derivatives appearing in the gauge transformations \eqref{fieldsgaugetransform} act on scalar quantities and are thus already covariant. 
The curvatures $H_{\mu \nu A}$ and $B_{\mu \nu}$, being two-forms, are independent of the metric, so that (\ref{defHmunuTmunu}) and (\ref{defBmunu}) are already covariant as well.
The same is not true for $\G_{\mu \nu \r}$, whose covariant definition reads
\ba
\label{Gamma:cov}
\Gamma_{\mu\nu\lambda} 
=
2\left[\dc_{(\mu}B_{\nu)\lambda}
- \frac{1}{2}\left( H_{\lambda \mu A} e_{\nu}^{\ A}+H_{\lambda \nu A} e_{\mu}^{\ A}\right)
\right]
\ .
\ea
It is easy to check that $B_{\m \n}$, $H_{\m \n A}$ and $\G_{\m \n \l}$ are gauge invariant also when the theory is coupled to a curved geometry. 

It is worth emphasizing the role played by the internal spacetime, crucial in our realization of the monopole-dipole-momentum algebra, by seeing what would happen if all the indices were external. In that case, we would need to introduce covariant derivatives into the dipole gauge transformation in (\ref{fieldsgaugetransform}) and in the definition of $H_{\m \n \l}$. Consequently, $H_{\m \n \l}$ and $\G_{\m \n \l}$, although diffeomorphism covariant, would not be gauge invariant.

To conclude this section, thanks to the internal space realization of the dipole symmetry, we are able to write a gauge invariant and diffeomorphism covariant fracton theory. The action is given by
\be
S = \int d^4 x  \sqrt{- \g} \,  \mc{L} \ ,
\ee
where $\g = \det ( \g_{\m \n})$ and $\mc{L}$ is given by \eqref{lag_quad_2}, external indices are contracted with the background metric $\g_{\mu\nu}$ and the curvatures are treated as we have just described in the present section: namely, we keep track of the internal index of $H_{\mu\nu A}$ and adopt \eqref{Gamma:cov} as the definition for $\Gamma_{\mu\nu\lambda}$.

\section{Interactions}
\label{secinteractions}

In the case of linearized gravity, the interaction of the massless spin-2 field with itself or other fields must be through a conserved symmetric tensor, which in a generic theory has to be the energy-momentum tensor. However, the energy-momentum tensor itself depends on additional interaction terms, so that doing this self-consistently gives as a result the full non-linear theory of gravity coupled to matter \cite{Kraichnan:1955zz,Feynman:1996kb,Gupta:1952zz,Deser:1969wk} (see also \cite{Ortin:2004ms} for a nice account of these facts). This argument, and Weinberg's soft graviton theorem \cite{Weinberg:1964ew} are seen as proof that two-derivative interactions of a massless spin-2 field are universal.

In principle the theory obtained from spacetime dipole symmetry enjoys a wider freedom since the conditions \eqref{eq:ward} are less constraining than energy-momentum conservation. For instance, a two-derivative symmetric tensor which is not conserved but satisfies the continuity equation is
\begin{equation}\label{eq:Jmn}
    J^{\mu\nu}=c_V\left(\partial^\mu V^\nu+\partial^\nu V^\mu-2\eta^{\mu\nu} \partial_\sigma V^\sigma\right) \ ,
\end{equation}
with $V^\mu$ being an arbitrary vector operator containing at most one derivative and $c_V$ an arbitrary constant. As a simple example, one could take $V^\mu=J^\mu$, the $U(1)$ global current of a complex scalar field
\begin{equation}\label{eq:currentV}
    J^\mu=\frac{i}{2}\left(\phi^*\partial^\mu \phi-\partial^\mu \phi^* \phi\right)\ .
\end{equation}
In this case
\begin{equation}
    J^{\mu\nu}=i\left(\phi^*\partial^\mu \partial^\nu\phi-\partial^\mu\partial^\nu \phi^* \phi\right)-2\eta^{\mu\nu}\partial_\sigma J^\sigma\ .
\end{equation}
Integrating by parts, one can rewrite the coupling to $V^\mu$ also as
\begin{equation}\label{eq:mattinter}
    \frac{1}{2}h_{\mu\nu}J^{\mu \nu}=-c_V H_\mu V^\mu+\text{total\, derivative}\ .
\end{equation}
However, in this form it is apparent that the spin-2 component of the symmetric field does not enter in the interaction. We encounter a similar situation when considering the self-coupling of the spacetime dipole fields. Gauge invariant interaction terms can be assembled using the invariants curvatures \eqref{eq:invcurvatures}, but if we restrict to two-derivative terms, we are forced to use $B_{\mu\nu}$ as one of the factors. Thus there is no gauge-invariant two-derivative cubic self-coupling for the symmetric field. 

A possible caveat to the last statement is that the cubic term needs to be invariant only up to a total derivative, so one may wonder if a cubic term for the symmetric field is still possible. This would require adding to the action a term of the form $h_{\mu \nu} J^{\mu\nu}$, with $J^{\mu\nu}$ quadratic in the symmetric field and containing two derivatives. Gauge invariance requires $\partial_\mu \partial_\nu J^{\mu\nu}=0$ off-shell. As we detail in appendix \ref{app:cubic}, there is no term of this kind.
This is similar to the non-linear generalization of the standard spin-2 theory. Full consistency of the interaction terms requires extending the linearized gauge trasformations to the full diffeomorphism trasformations. The lowest order correction corresponds to a spacetime translation and makes the transformation field-dependent:
\begin{eqnarray}\label{fie:dep}
    \delta h_{\mu\nu}=\partial_{\mu}\xi_\nu+\chi\,\xi^\alpha \partial_\alpha h_{\mu\nu}\ ,
\end{eqnarray}
where $\chi$ is the expansion parameter proportional to Newton's constant. In principle, a similar approach could be considered for our theory. However, since this involves incorporating field-dependent terms in the gauge transformations, it exceeds the scope of the present paper.

An alternative way to produce self-interactions is through spontaneous symmetry breaking of the dipole symmetry. For this, we can generalize the matter fields introduced in \cite{Afxonidis:2023tup} to the Lorentz covariant theory. In particular, we can take a set of complex scalar fields $\phi_a$, $a=0,1,2,3$ with dipole charges $d_a^A=d\,\delta_a^A$ (recall that each dipole charge is a vector in the internal space). Under a gauge transformation each of the fields transforms by a phase
\begin{equation}
    \phi_a\quad\longrightarrow\quad e^{i\lambda_{1\,A} d_a^A}\phi_a \ , 
\end{equation}
where there is no sum over $a$.
The covariant derivatives of these fields are simply
\begin{equation}
    D_\mu \phi_a=\partial_\mu \phi_a-i b_{\mu A}d_a^A \phi_a\ .
\end{equation}
This allows to add new invariant interaction terms to the action, like for instance
\begin{equation}
    \Gamma_{\mu\nu}^{\phantom{\mu\nu}\sigma}\Gamma^{\mu\nu\rho}\sum_a \eta_{_{BC}}\, e_\sigma^{_B} d_a^{_C} \left(i\phi_a^* D_\rho \phi_a-iD_\rho\phi_a^*  \phi_a\right) \ .
\end{equation}
If the scalar fields acquire an expectation value $\phi_a=v$, this term introduces a cubic interaction for the dipole gauge field
\begin{equation}
   \sim  2d\,v^2\,\Gamma_{\mu\nu}^{\phantom{\mu\nu}\sigma}\Gamma^{\mu\nu\rho} b_{\sigma\rho} \ .
\end{equation}
In terms of the symmetric and antisymmetric fields, this would introduce new interactions, including in principle cubic self-interactions for the symmetric field. It would be interesting to explore if those reproduce or not the low-energy scattering amplitudes of the graviton in the usual theory, although this might require introducing exotic kinetic terms for the dipole fields, in order to avoid giving a mass to the spin-2 component of the symmetric field. 

\subsection{Dipole conservation and fractons}

The covariant version of the fractonic conservation equations \eqref{eq:ward} and \eqref{eq:wardf} does not result in an actual conservation of the dipole moment of the charge density, since
\begin{equation}
    \partial_t Q^i\equiv \partial_t\int d^3x\, x^i J^t=\partial_t \int d^3 x\, J^{ti} \ .
\end{equation}
However, if we identify $J^{ti}$ as an intrinsic dipole moment, then the following combination is conserved
\begin{equation}
    Q_{\text{tot}}^i=\int d^3 x\left(x^i J^t-J^{ti}\right)\ .
\end{equation}
In addition, the time component of the dipole charge is also conserved
\begin{eqnarray}
    Q^t_{\text{tot}}=\int d^3 x\left( t J^t-J^{tt}\right) \ .
\end{eqnarray}

This actually generalizes to a covariant version of the conservation of dipole charge. Given a codimension-one hyperplane $\Sigma$, which could be space- or time-like, we pick an orthonormal basis of constant vectors $\{\hat{n},e^a \}$, with $\hat{n}$ the unit normal and $e^a$, $a=1,2,3$, vectors spanning the directions along the hyperplane. Then, the `$\Sigma$-dipole' charge can be defined as
\begin{eqnarray}
    Q_\Sigma^a=\int_\Sigma d^3x \, e^a_\mu \hat{n}_\nu\left( x^\mu J^\nu-J^{\nu\mu} \right) \ .
\end{eqnarray}
This is `conserved' in the sense that it has the same value on hyperplanes parallel to $\Sigma$:
\begin{equation}
    \hat{n}^\mu \partial_\mu Q_\Sigma^a=0 \ .
\end{equation}
When $\Sigma$ is just a spatial slice then this becomes the spatial dipole conservation we wrote above. One can easily check that the component of the charge transverse to the hyperplane is also conserved in the sense above
\begin{eqnarray}
     Q_\Sigma^n=\int_\Sigma d^3x \, \hat{n}_\mu \hat{n}_\nu\left( x^\mu J^\nu-J^{\nu\mu} \right) \ , \q \q   \hat{n}^\mu \partial_\mu Q_\Sigma^n=0\,.
\end{eqnarray}

One might wonder what kind of configuration is sourced by a ``fracton'', a point-like charge
\begin{equation}
    J^t=q\, \delta^{(3)}(\bm{x})\ .
\end{equation}
This can be obtained from a current with only non-zero component $J^{it}=J^{ti}$, such that $J^t=\partial_i J^{it}$. The form of this current is
\begin{eqnarray}
    J^{ti}=J^{it}=-\frac{q}{4\pi}\partial^i\frac{1}{r} =\frac{q}{4\pi}\frac{x^i}{r^3}  \ .
\end{eqnarray}
Then, introducing this current in \eqref{eq:eqh}, a configuration sourced by a fracton (in the absence of other matter fields) is
\begin{equation}
    h_{it}=\frac{q}{16\pi g_2}\partial_i r=\frac{q}{16\pi g_2} \frac{x_i}{r} \ .
\end{equation}
Thus it is a hedgehog configuration that in linearized gravity would be produced by a large diffeomorphism $\xi_t=\frac{q}{16\pi g_2} r$.

Although the fracton carries zero spatial dipole charge, it has a non-zero spacetime dipole charge. Take the hyperplane $\Sigma_i$ located at $x^i=x_0^i\neq 0$ for a fixed value of $i$. Then, 
\begin{eqnarray}
    Q^t_{\Sigma_i}=\int dt \int_\parallel d^2x \left( t J^i-J^{it}\right)=-\frac{q}{4\pi}x_0^i \int dt \int_\parallel d^2 x \frac{1}{r^3}\Big|_{x^i=x_0^i} \ .
\end{eqnarray}
Here the symbol $\parallel$ refers to the two spatial directions along the hyperplane.
The result is
\begin{eqnarray}
    Q^t_{\Sigma_i}=-\frac{q}{2}\operatorname{sign}(x_0^i)\int dt \ .
\end{eqnarray}
Therefore, from the point of view of the full set of `conserved' quantities, the fracton is actually a spacetime dipole in the sense defined above. An intuitive explanation is that the fracton has a worldline extended in time, which is along a direction contained in the hyperplanes $\Sigma_i$. This suggests that in order for all dipole charges to vanish, the `true' fracton should be some type of instanton localized at a single spacetime point.

\subsection{Map to linearized gravity solutions}
\label{subsecmaptolinearized}

Examining the equation for the symmetric field \eqref{eq:eqh}, one immediately sees that it is possible to map linearized gravity solutions to solutions of the spacetime dipole theory coupled to matter through a current of the form \eqref{eq:Jmn}. The map corresponds to finding solutions to the equations
\begin{equation}
    G^{\mu\nu}=0\ ,\qquad V^\mu=\frac{2 g_2}{c_V} H ^\mu\ .
\end{equation}
Let us study a class of solutions of linearized gravity that are also solutions to the full Einstein equations in the Kerr-Schild form of the metric. In the absence of matter, this includes plane wave solutions as well as Schwarzschild and Kerr solutions. 

The full metric in the Kerr-Schild form is $g_{\mu\nu}=\eta_{\mu\nu}+h_{\mu\nu}$, with
\begin{equation}
    h_{\mu \nu}=\Phi \,\ell_\mu \ell_\nu\ ,
\end{equation}
and the null vector $\ell_\mu$ satisfies 
\begin{equation}
   \ell^\mu \ell_\mu=0 \ ,\qquad
   \ell^\mu \partial_\mu \ell_\nu=S\, \ell_\nu\ ,
\end{equation}
where $S$ is a scalar function and indices are raised with the Minkowski metric $\eta^{\mu\nu}$. The vector $H_\mu$ is thereby proportional to the null vector,
\begin{equation}\label{H:null}
H_\mu=\partial^\sigma h_{\sigma \mu}-\partial_\mu h=\partial_\sigma(V \ell^\sigma \ell_\mu)=[\partial_\sigma(V \ell^\sigma)+S V]\, \ell_\mu\ .
\end{equation}
Thus $H_\mu H^\mu=0$. In addition, since vacuum Einstein's equations are satisfied, the divergence vanishes,
\begin{equation}\label{div:H}
    \partial_\mu H^\mu\propto \eta_{\mu\nu}G^{\mu\nu}=0\ .
\end{equation}

In order to construct the map between linearized gravity solutions and solutions of the spacetime dipole theory, we add to the latter a matter sector constituted by a complex scalar field coupled to an Abelian gauge field
\begin{equation}\label{matter}
{\cal L}_{\text{matter}}=-\frac{1}{4}F_{\mu\nu}F^{\mu\nu}-D_\mu\phi^*D^\mu \phi-\frac{\lambda}{2}(|\phi|^2-v^2)^2\ ,
\end{equation}
where the covariant derivative is
\begin{equation}
D_\mu\phi=(\partial_\mu-iq A_\mu)\phi\ .
\end{equation}
We have introduced a potential triggering symmetry breaking, the map also works if there is no potential, \emph{i.e.} for $\lambda=0$. We now couple the fields in the matter sector to the spacetime dipole fields through the interaction terms
\begin{equation}
{\cal L}_{\text{int}}=-c_V H_\mu J^\mu-c_{BF} B_{\mu\nu} F^{\mu\nu}\ ,
\end{equation}
where $J^\mu$ is the current
\begin{equation}
J^\mu=\frac{i}{2}\left(\phi^*D^\mu\phi-D^\mu\phi^*\phi\right)\ .
\end{equation}
Assuming the action for $B_{\mu\nu}$ is \eqref{eq:Baction}, the equations of motion for $B_{\mu\nu}$, the gauge field $A_\mu$ and the scalar $\phi$ are respectively
\begin{subequations}
\begin{eqnarray}
\frac{g_1-2g_2}{3}\, \partial_\sigma {\cal H}^{\sigma \mu\nu}-\frac{\alpha_2}{2} B^{\mu\nu}-c_{BF}F^{\mu\nu}&=& 0\ ,\\
\partial_\sigma F^{\sigma\mu}+2c_{BF}\, \partial_\sigma B^{\sigma \mu}-2q  J^\mu
-q c_V |\phi|^2H^\mu &=&0\ ,\\
D_\mu D^\mu\phi-ic_V H^\mu D_\mu \phi-\frac{i}{2} c_V \partial_\mu H^\mu \phi-\lambda(|\phi|^2-v^2)\phi & =& 0\ ,
\end{eqnarray}
\end{subequations}
where ${\cal H}^{\sigma\mu\nu}$ has been defined in \eqref{cal_H}.
Imposing the condition
\begin{equation}
\alpha_2=4c_{BF}^2\ ,
\end{equation}
we find the solutions
\begin{equation}\label{sol:map}
\phi=v\ ,\qquad 
A_\mu =-\frac{c_V}{2q}H_\mu\ ,\qquad  B_{\mu\nu}=-\frac{1}{2c_{BF}}F_{\mu\nu}\ ,
\end{equation}
so that $A_\mu$ is null and divergenless according to \eqref{H:null} and \eqref{div:H}.
The equations in \eqref{sol:map} map the most general Kerr-Schild metric which solves the vacuum Einstein's equations to solutions of the particular spacetime dipole theory with matter content specified in \eqref{matter}. 

In particular cases, it is possible to map specific Kerr-Schild solutions to solutions of a spacetime dipole theory with a reduced matter content compared to \eqref{matter}. 
Let us examine in more detail two such cases: the plane-wave solutions, where no matter is needed, and Schwarzschild solutions, where the map can be made adding just a massless scalar field.

\paragraph{\bf \large Plane waves}\ \newline

\noindent
For plane-wave solutions one may select $z$ as the direction of propagation of the wave
\begin{equation}
\Phi=\varphi(t-z,x,y)\ , \qquad
\ell_\mu dx^\mu=-dt+dz\ .
\end{equation}
Note that 
\begin{equation}
    \ell^\mu\partial_\mu \Phi=0\ ,\qquad \partial_\mu \ell^\mu=0\ ,
\end{equation}
hence $\ell_\mu$ is a Killing vector.
Solutions for the amplitude must be harmonic functions in the $(x,y)$ plane:
\begin{equation}
    (\partial_x^2+\partial_y^2) \varphi =0\  .
\end{equation}
A straightforward calculation of \eqref{eq:HG}, shows that for plane-wave solutions $H_\mu=0$, so that these are also solutions of the spacetime dipole theory without matter.

Particular cases are shock waves
\begin{equation}
    \varphi(t-z,x,y)=A_0\, \delta(t-z)\, \log\left(\frac{x^2+y^2}{R^2}\right) \ ,
\end{equation}
which are actually produced by a highly-energetic particle moving at the speed of light, but are vacuum solutions outside the particle's trajectory. Truly vacuum solutions are exact waves
\begin{equation}
    \varphi(t-z,x,y)=A_+(t-z) (x^2-y^2+2i xy)+A_-(t-z)(x^2-y^2-2i x y)\ .
\end{equation}
More generally, the solutions are a sum of a holomorphic function on the $(x,y)$ plane and its conjugate
\begin{equation}
    \varphi(t-z,x,y)=A(t-z,x+iy)+\bar{A}(t-z,x-iy)
\end{equation}

\paragraph{\bf \large Schwarzschild solution}\ \newline

\noindent
The Schwarzschild solution in Kerr-Schild form is
\begin{equation}
\Phi=\frac{2m}{r}\ ,\qquad
\ell_\mu dx^\mu=-dt-\frac{x^i dx^i}{r}\ ,\qquad  r^2=\sum_{i=1}^3 (x^i)^2\ .
\end{equation}
The null vector $\ell_\mu$ satisfies the identities 
\begin{equation}
 \partial_\mu \ell^\mu=-\frac{2}{r}\ ,\qquad \ell^\mu \partial_\mu r=-1\ .
\end{equation}
Then, from the definition in \eqref{eq:HG},
\begin{equation}
    H_\mu=-\left( \Phi'+\frac{2}{r}\Phi\right)\ell_\mu=-\frac{2m}{r^2} \ell_\mu\ .
\end{equation}
When $V^\mu=J^\mu$, the current in \eqref{eq:currentV}, there is a solution which is an spherical wave of a massless field
\begin{equation}
    \phi(x)=A\frac{e^{\mp i\omega (t+r)}}{r}\ ,\qquad \partial^2 \phi=0\ .
\end{equation}
The condition on the energy of the wave is
\begin{equation}
   E= \omega|A|^2=\pm\frac{4 m g_2}{c_V}\ .
\end{equation}
The sign will be chosen according to the sign of $g_2$. Note that if we add to the action of a massless scalar the coupling \eqref{eq:mattinter}, the conserved $U(1)$ current would not be just given by \eqref{eq:currentV}, but there will be an additional term
\begin{equation}
    J_{\rm tot}^\mu= J^\mu+\frac{c_V}{2} H_\mu \phi^*\phi\ .
\end{equation}
Similarly, the equations of motion will be modified
\begin{equation}
    \partial^2 \phi-i c_V H^\mu \partial_\mu \phi-\frac{i}{2}c_V \partial_\mu H^\mu \phi=0\ .
\end{equation}
The spherical wave is not a solution to the modified equations, but one could in principle use a perturbative expansion in $c_V$ to systematically find corrections. Alternatively, one could add a term  to the scalar action
\begin{equation}
\Delta {\cal L}=\frac{c_V^2}{4g_2}J_\mu J^\mu\ .
\end{equation}
Adding this term actually corresponds to completing the square
\begin{equation}
    {\cal L}_{H^2}=g_2\left( H_\mu-\frac{c_V}{2 g_2}J_\mu\right)^2\ .
\end{equation}
This cancels the contributions proportional to $H^\mu$ in the equations of motion of the scalar field and in $J^\mu_{\rm tot}$, in such a way that the scalar spherical wave and Schwarzschild metric are exact solutions of the coupled scalar field and spacetime dipole theory.

\section{Conclusions}

We studied the Lorentz-covariant generalization of gauge theories where both an Abelian charge and its  \emph{spacetime} ``dipole moment'' are conserved. In general, the resulting theory contains a massive gauge-invariant antisymmetric field and a massless symmetric field. This latter transforms under gauge transformations as a ``fracton'' generalization of gravity. Given the reduced amount of symmetry, the action allows extra terms with respect to standard linearized gravity. For the action of the symmetric field there are two possible terms with independent coefficients $g_1$, corresponding to the action of linearized gravity, and $g_2$, corresponding to a new term forbidden by diffeomorphism invariance but allowed by the reduced symmetry. Such point was already discussed in \cite{Blasi:2022mbl} but the realization proposed in the present paper allows for an additional mixing term with the antisymmetric field. 

If $g_2=0$ the action becomes that of linearized gravity, but even for $g_2\neq0$ we have constructed a map between solutions of linearized gravity and the Lorentz covariant dipole theory coupled to matter. The solutions are in Kerr-Schild form and in some cases they are also solutions of the full Einstein equations. In this map, the components of the symmetric current equal the linearized Einstein tensor
\begin{equation}
    J^{(\mu\nu)}=-4(g_1-g_2) G^{\mu\nu} \ .
\end{equation}
For instance, in the case of a Schwarzschild black hole, the only non-zero component is $J^{tt} =64\pi m(g_1-g_2)\delta^{(3)}(\bm{x})$. This translates into having a point-like dipole (plus additional gauge and/or scalar fields). It should be noted that in this case the spacetime dipole charge corresponds to a vector pointing in the time direction, so it is not a spatial dipole in the usual sense.

We have also revisited the question concerning the gauge fixing, which in our formalism can be treated with standard tools. In particular we used the BRST approach and show why scalar and vector gauge fixings in the classification of \cite{Blasi:2022mbl,Bertolini:2023juh} are both possible. This corresponds to having both a scalar and a vector-like gauge symmetry stemming from the charge and the dipole transformations, respectively.

The fractonic gauge transformation $\delta h_{\mu\nu} = - \partial_\mu \partial_\nu \lambda_0$ encountered in \eqref{eq:gaugetrh} corresponds to a standard linearized diffeomorphism $\delta h_{\mu\nu} = 2\partial_{(\mu} \xi_{\nu)}$ where the vector parameter $\xi_\mu$ is longitudinal $\xi_\mu=-\partial_\mu\lambda_0$. We are therefore concerned with a complementary case to the ``transverse diffeomorphisms", $\partial_\mu \xi^\mu=0$, corresponding to linearized unimodular gravity \cite{Buchmller1988EinsteinGF,Kreuzer1990GaugeTO,lvarez2005CanOT,Alvarez:2006uu}, where the determinant of the full metric --the trace of $h_{\mu\nu}$ at the linearized level-- is invariant by construction.
It is intriguing to observe that asking the trace of the symmetric field to be invariant also in our case implies $\Box \lambda_0 =  0$ 
that corresponds to the scalar gauge fixing \eqref{eq:gaugefixscalar} for $\xi\to 0$. This condition is compatible with global dipole transformations $\lambda_{1\,\mu}=b_\mu$, $\lambda_0=-b_\mu x^\mu+c$, which are by definition those which leave the gauge potentials invariant \cite{Caddeo:2022ibe}. A related connection was made in \cite{Gromov:2017vir}, where invariance under area-preserving spatial diffeomorphisms was imposed to construct a theory dual to topological elasticity and avoid the issues of coupling the tensor theory to a background geometry.

Another advantage of realizing the dipole symmetry with ordinary one-form gauge fields is that it allows one to couple the theory to a curved background geometry without spoiling the dipole gauge symmetry. This thus avoids the issues previously discussed in the literature concerning the coupling of a rank-two gauge field to a generic curved background \cite{Gromov:2017vir,Slagle:2018kqf,Jain:2021ibh, Bidussi:2021nmp}.

In addition, we have explored the physical spectrum of the free theory. For generic values of the couplings, the symmetric field has physical massless modes with helicities $\pm 2$, $\pm 1$ and $0$. We have considered cubic self-interactions and interactions with matter. The two-derivative interactions we were able to construct involve only the spin-$0$ and spin-$1$ components of the symmetric field. This is somewhat counter-intuitive on the basis that a smaller gauge symmetry should in principle allow more freedom. However, the massless spin-$2$ part remains decoupled as it occurs for an ordinary spin-$2$ field. 

In order to have non-trivial two-derivative interactions, it is necessary in the usual gravity theory to modify the gauge transformation of the spin-$2$ field as in \eqref{fie:dep}. Possibly a similar modification may be also necessary in the present case. Along these same lines, it would be interesting to investigate whether Deser's all-order approach \cite{Deser:1969wk} could be generalized to our case.

A different avenue of future research would be to introduce additional matter fields inducing self-interactions of the spin-2 component of the symmetric field, in particular this could be done through spontaneous breaking of the dipole symmetry as commented in Section \ref{secinteractions}. It is interesting to explore whether this possibility can be pursued without the spin-2 component becoming massive.

Another natural extension of the present analysis consists in considering higher-multipole gauge symmetries. It is reasonable to expect that this could yield non-standard interactions for a spin-$2$ massless component, in a similar way as the dipole symmetry does for the spin-1 and spin-0 components. On top of this, it would be quite interesting \emph{per se} to study higher-multipole symmetries as candidates for alternative formulations of higher-spin theories.

A pertinent question in modified theories of gravity is stability in the absence of the full diffeomorphism invariance. Relying on the mode decomposition of the fields and the equations of motion, we expect the model studied in this paper to be stable for some suitable region in the space of couplings $g_1$ and $g_2$. The contributions of the individual modes to the energy density is semi-positive definite, see Appendix \ref{app:eoms}. The semi-positive character is particularly interesting because it arises from a null contribution to the energy density from the vector sector, an aspect which calls for further investigation. An exhaustive discussion of classical stability requires a systematic study of the canonical structure and constraints. It constitutes a necessary step in view of quantization, and we postpone it to the future.

As a final comment, the background $e_\m ^{\ A}$ enters the action of our model in a way that explicitly
breaks the internal transformations generated by $\xi^A$ \eqref{fieldsgaugetransform}. 
Yet, one could generalize the model by considering additional Nambu-Goldstone fields that shift under $\xi^A$, thereby investigating possible spacetime generalizations of the dual elastic theory of \cite{Caddeo:2022ibe}. In this context, it would be interesting to relate the physical scale $M$ to a dynamical symmetry-breaking scale.
Such a breaking would bridge internal and external spacetimes by locking internal and external transformations, in a similar spirit to the framework for low-energy effective field theories discussed in \cite{Nicolis:2015sra}. In particular, the temporal part of such a breaking could entail a preferred time foliation similar to the khronon scenario and Lifshitz gravity \cite{Horava:2009uw,Blas:2010hb,Creminelli:2012xb}.

\section*{Acknowledgements}

We would like to thank Erica Bertolini, Alberto Blasi, Matteo Carrega, Antón Faedo, Silvia Fasce and Nicola Maggiore for discussions and feedback. The work of A.C. is supported by Ministerio de Ciencia e Innovación de
España under the program Juan de la
Cierva-formación. This work is partially supported by the AEI
and the MCIU through the Spanish grant PID2021-123021NB-I00 and by FICYT
through the Asturian grant SV-PA-21-AYUD/2021/52177.

\appendix

\section{Equations of motion}
\label{app:eoms}

The symmetric equation of motion descending from \eqref{lag_quad_2}
is
\ba
\label{eq:heq}
    0&=& \left(\alpha_1+3\alpha_3+\frac{3\alpha_5}{2}\right) \partial^2 h^{\nu\lambda} 
- \frac{1}{2} \left(\alpha_1 + 3\alpha_3 - 2\alpha_4 + \frac{3\alpha_5-\a_{6} - \a_{7}}{2}\right) \partial_\mu(\partial^\nu h^{\mu\lambda}+\partial^\lambda h^{\mu\nu}) \nb \\ \nonumber
&-&\left(2\alpha_4  +\frac{1}{2}(\alpha_6+\alpha_7) \right)\left(\partial^\nu\partial^\lambda h
- \eta^{\nu\lambda}\partial^2 h
+ \eta^{\nu\lambda}\partial_\sigma\partial_\rho h^{\sigma\rho}\right)  \\
&+&\frac{1}{2}\left(-\alpha_1+3\alpha_3+2\alpha_4 -\frac{1}{2}\alpha_7 \right)\partial_\mu(\partial^\nu B^{\mu\lambda}+\partial^\lambda B^{\mu\nu})\ .
\ea 
Adopting the definitions introduced in \eqref{eq:HG} and 
\begin{equation}\label{long_B}
    B^\mu \equiv \partial_\sigma B^{\sigma\mu}\ ,
\end{equation}
which automatically satisfies $\partial_\mu B^\mu=0$,
equation \eqref{eq:heq} can be conveniently re-written as 
\ba
\label{eq:symmG}
   0&=& \ \beta\, G^{\nu\lambda}
   +\frac{1}{2}\left(\beta+2\alpha_4 +\frac{1}{2}(\alpha_6+\alpha_7) \right)\left(\partial^\nu H^\lambda+\partial^\lambda H^\nu-2 \eta^{\nu\lambda}\partial_\sigma H^\sigma \right) + \nb \\
   &+&
   \frac{1}{2}\left(-\alpha_1+3\alpha_3+2\alpha_4 -\frac{1}{2}\alpha_7 \right)(\partial^\nu B^\lambda+\partial^\lambda B^\nu)\ ,
\ea
where $\beta$ is defined in \eqref{beta}. Assuming the decoupling condition \eqref{eq:decoupling}, the trace and the divergence of \eqref{eq:symmG} yield respectively
\begin{align}\label{lon_H}
    \big[2 \beta+(d-1) \pr{4 \alpha_4  + \alpha_6+\alpha_7}\big]\partial_\sigma H^\sigma &=0\ ,\\
    \label{H2}
    \left(2 \beta+ 4 \alpha_4  +  \alpha_6+\alpha_7 \right)\left(\partial^2 H^\lambda- \partial^\lambda\partial_\sigma H^\sigma\right)&=0\ .
\end{align}
For $\beta\neq -2\alpha_4  -\frac{1}{2}(\alpha_6+\alpha_7)$ and $\beta\neq -\frac{1}{2}(d-1)(4 \a_{4 }+\alpha_6+\alpha_7)$, combining \eqref{lon_H} and \eqref{H2}, we get
\begin{equation}
    \label{div_acc}
    \partial^2 H^\mu=0\ ,\qquad 
    \partial_\mu H^\mu=0\ .
\end{equation}

The antisymmetric equation of motion descending from \eqref{lag_quad_2}
is
\ba
\label{eq:Beq}
    0 &=& \left(\alpha_1+\alpha_3-\frac{\alpha_5}{2}\right) \partial^2 B^{\nu\lambda}
- \frac{1}{2}\left(\alpha_1 - \alpha_3 - 2\alpha_4 + \frac{\alpha_5 + \a_{6} - \a_{7}}{2}\right)\partial_\mu(\partial^\nu B^{\mu\lambda}-\partial^\lambda B^{\mu\nu}) + \nb \\
&-&2\alpha_2 B^{\nu\lambda}  
 +\frac{1}{2}\left(-\alpha_1+3\alpha_3+2\alpha_4 - \frac{1}{2}\alpha_7 \right)\partial_\mu(\partial^\nu h^{\mu\lambda}-\partial^\lambda h^{\mu\nu})\ .
\ea
Again considering the decoupling condition \eqref{eq:decoupling}, the divergence of the antisymmetric equation \eqref{eq:Beq} yields
\begin{equation}
0=\left(\alpha_1+3\alpha_3+2\alpha_4-\frac{3 \alpha_5 + \a_6 - \a_7}{2}\right)\partial^2 B^{\lambda} -4\alpha_2 B^{\lambda}\ ,
\end{equation}
where we have recalled the definition of the longitudinal field $B^\mu$ given in \eqref{long_B}.
For the transverse part $\partial_\mu B^{\mu\nu}_\perp=0$, we have instead
\begin{equation}
0=\left(\alpha_1+\alpha_3-\frac{\alpha_5}{2}\right)\partial^2 B_\perp^{\nu\lambda} -2\alpha_2 B_\perp^{\nu\lambda} \ .   
\end{equation}

\subsection{Constraints on the mode expansion from the equations of motion}
\label{con:mot}

Consider the Fourier transform of the gauge-invariant vector introduced in \eqref{eq:HG} expressed in terms of the modes \eqref{spli_acc},
\begin{equation}
    \widetilde{H}^\mu 
    = ik_\sigma \varepsilon^{\sigma\mu}-i k^\mu \varepsilon^\mu_{\ \mu}
    = -2 C_1 i q^\mu + \omega\, C_{2\,\sigma}\epsilon_\sigma^\mu+\left(1-\frac{d}{2} \right)(C_3+2B) i k^\mu\ .
\end{equation}
Then, from the equation $\partial_\mu H^\mu=0=ik_\mu \widetilde{H}^\mu$ obtained in \eqref{div_acc}, we fix
\begin{equation}
    C_1=0\ .    
\end{equation}
Next, let us compute
\begin{equation}
    \label{sym_acc}
    \partial^\mu H^\nu+\partial^\nu H^\mu\to ik^\mu \widetilde{H}^\nu+ik^\nu \widetilde{H}^\mu=2\omega C_{2\, \sigma} ik^{(\mu}\epsilon_\sigma^{\nu)}+(d-2)(C_3+2B)k^\mu k^\nu\,.
\end{equation}
From this we obtain for the Fourier transform of the linearized Einstein tensor
\begin{equation}
    \label{ein_acc}
    \widetilde{G}^{\mu\nu}=k^\mu k^\nu \varepsilon^\lambda_{\ \lambda}-\left(ik^\mu \widetilde{H}^\nu+ik^\nu \widetilde{H}^\mu\right)=-2\omega C_{2\, \sigma} ik^{(\mu}\epsilon_\sigma^{\nu)}-(d-2)(C_3+B)k^\mu k^\nu \ .
\end{equation}
Then, from the equation of motion \eqref{eq:symmG}, we get
\ba
    \label{equ_acc}
    0&=&\omega\, C_{2\, \sigma}i k^{(\mu}\epsilon_\sigma^{\nu)}\left(2\alpha_4-\beta +\frac{1}{2}(\alpha_6+\alpha_7)\right)\\ \nonumber
    &+&\frac{1}{2}(d-2) k^\mu k^\nu \left[\left(2\alpha_4-\beta +\frac{1}{2}(\alpha_6+\alpha_7)\right)C_3+\left(4\alpha_4+\alpha_6+\alpha_7 \right) B \right]\ .
\ea
Assuming $\beta\neq 2\alpha_4  +\frac{1}{2}(\alpha_6+\alpha_7)$, we get the conditions
\begin{equation}
    \label{con_acc}
    C_{2\,\sigma}=0 \ , \quad \quad \sigma=1,2\ ,
\end{equation}
and
\begin{equation}
C_3=\frac{4\alpha_4  +\alpha_6+\alpha_7}{\beta-2\alpha_4 +\frac{1}{2}(\alpha_6+\alpha_7)}B\ .
\end{equation}

\subsection{Comments on stability}

In Subsection \ref{con:mot} we study how the equations of motion restrict the mode expansion. Enforcing such restrictions in \eqref{mod_exp} and \eqref{spli_acc}, we are left with
\ba
\label{eq:solhmn}
    \varepsilon^{\mu\nu}&=&A_\sigma \epsilon_\sigma^\mu\epsilon_\sigma^\nu
    +
    A_{3a} \frac{i k^{(\mu}\epsilon_a^{\nu)}}{\omega}+A_4 \frac{k^\mu k^\nu}{\omega^2}\\ \nonumber
    &+&B\left( \frac{\beta+2\alpha_4  +\frac{3}{2}(\alpha_6+\alpha_7)}{\beta-2\alpha_4  +\frac{1}{2}(\alpha_6+\alpha_7)}\eta^{\mu\nu}+\frac{\beta+  6 \alpha_4   +\frac{5}{2}(\alpha_6+\alpha_7)}{\beta-2\alpha_4  +\frac{1}{2}(\alpha_6+\alpha_7)}\frac{i k^{(\mu}q^{\nu)}}{\omega^2}\right)\ ,
\ea
where $A_4$ corresponds to the pure-gauge, unphysical part and both $\sigma$ and $a$ run over $\{1,2\}$. Computing the energy density, one finds that the contribution from the modes of spatial momentum $\boldsymbol{k}$ is given by
\begin{equation}\label{con:ene}
   c_2 \left(\tilde{A}_{-1}(\boldsymbol{k}) \tilde{A}_{-1}^*(\boldsymbol{k})
   +
   \tilde{A}_{1}(\boldsymbol{k}) \tilde{A}_{1}^*(\boldsymbol{k})\right)
   + c_0\,
   \tilde{B}(\boldsymbol{k}) \tilde{B}^*(\boldsymbol{k})\ ,
\end{equation}
where
\begin{equation}
    c_0 = \omega\, \frac{2 g_2 \left(2 g_1^2 - g_1 g_2 - g_2^2\right)}{\pi  (g_2-2 g_1)^2}\ ,\qquad
    c_2 = \omega\, \frac{g_1-g_2}{\pi }\ ,
\end{equation}
with $k\equiv |\boldsymbol{k}|$.
One can get a positive definite contribution for suitable choices for the coefficients $g_1$ and $g_2$. For instance, assuming $g_1>0$ and small $g_2$, namely $|g_2|<g_1$, one has a positive definite contribution \eqref{con:ene} for $g_2>0$. The other possibility is $g_2<0$ and $|g_2|>2 g_1$. Note that the vector modes associated to the coefficients $A_{3a}$ do not enter in \eqref{con:ene}, thus they do not contribute to the energy density.

\section{Computational details}

In this appendix we collect some useful formul$\ae$ or intermediate passages for the computations described in Subsection \ref{pro:sec} of the main text.

In order to express the equations of motion for the symmetric field \eqref{eq:eqh} in the Fourier form \eqref{eq:eqh:fou}, we adopt the definitions given in \eqref{ten:def} and we have the following intermediate results:
\begin{align}
    G^{\mu\nu}
    & \rightarrow 
    \left(
    -\frac{1}{2} k^2 I^{\mu\nu,\alpha\beta}
    +\frac{1}{2} K_1^{\mu\nu,\alpha\beta}
    + K_2^{\mu\nu,\alpha\beta}
    \right)\varepsilon_{\alpha\beta} \ ,
    \\ 
    \partial^\mu H^\nu + \partial^\nu H^\mu - 2 \eta^{\mu\nu} \partial_\sigma H^\sigma 
    &\rightarrow
    -\left( 
    \frac{1}{2} K_1^{\mu\nu,\alpha\beta}
    + 2 K_2^{\mu\nu,\alpha\beta}
    \right)\varepsilon_{\alpha\beta} \ .
\end{align}

In what follows, we collect some technical details useful to the purpose of getting the propagator \eqref{prop}, or also to just check \eqref{green}. Let us define
\begin{equation}
    K_3^{\mu\nu,\alpha\beta} \equiv k^\mu k^\nu k^\alpha k^\beta\ ,
\end{equation}
and consider it alongside the definitions given in \eqref{ten:def}. Besides, it is convenient to introduce the compact notation
\begin{eqnarray}\label{not:pro}
    A\cdot B \equiv A^{\mu\nu,\rho\sigma} B_{\rho\sigma,\alpha\beta}\ ,
\end{eqnarray}
and, since we focus only on tensors possessing the symmetry $A^{\mu\nu,\alpha\beta}=A^{\alpha\beta,\mu\nu}$, we have $A\cdot B = B\cdot A$, up to straightforward lowering/raising of indexes that we leave implicit. The set of tensors ${\cal B} \equiv \{I,K_1,K_2,K_3\}$ is closed under the product $\cdot$ and in particular we have
\begin{align}\label{dot:bas}
\begin{split}
    I\cdot I &= 2 I\ , 
    \qquad 
    I\cdot K_1 = 2 K_1\ ,
    \qquad 
    I\cdot K_2 = 2 K_2\ ,
    \qquad 
    I\cdot K_3 = 2 K_3\ , 
    \\ 
    K_1\cdot K_1 &= 2k^2 K_1 + 8 K_3\ ,
    \qquad 
    K_1\cdot K_2 = -4 K_3\ ,
    \qquad 
    K_1\cdot K_3 = 4k^2 K_3\ ,
    \\ 
    K_2\cdot K_2 &= 3k^2 K_2 + 4 K_3\ ,
    \qquad 
    K_2 \cdot K_3 = -k^2 K_3\ ,
    \\ 
    K_3\cdot K_3 &= (k^2)^2 K_3\ .
\end{split}
\end{align}
In the first line of \eqref{dot:bas} we just have specific cases of the general relation
\begin{eqnarray}
    I\cdot A = 2 A\ ,
\end{eqnarray}
valid for all $A \in \text{span}( {\cal B})$, which amounts to recalling that $I$ is proportional to the identity with respect to the product \eqref{not:pro}.

\section{Cubic couplings}
\label{app:cubic}

In this appendix we detail on the search for graviton cubic self-couplings that are gauge-invariant up to total derivatives. We look for terms of the form
\be
\label{selfcouplingtotder}
\mc{L}_{h J_{h}} = h_{\m \n} J_{h} ^{\m \n} \ ,
\ee
where $J_{h} ^{\m \n}$ is a symmetric rank-two tensor quadratic in $h_{\m \n}$ that features two derivatives. In order for the cubic term to be invariant under the gauge transformation \eqref{eq:gaugetrh}, the current $J_h^{\mu\nu}$ should be gauge invariant and doubly-conserved off-shell, namely 
\be
\d J_{h} ^{\m \n}=0 \ , \q \q \q  \p_{\m} \p_{\n} J_{h}^{\m \n} = 0 \q \q \q \text{off-shell} \ .
\ee
Before imposing gauge invariance and conservation, the most general $J_{h}^{\m \n}$ reads (see e.g. \cite{Ortin:2004ms})
\ba
J_{h}^{\m \n} &=& c_{1} \p^{\m} h_{\l \r} \p^{\n} h^{\l \r} + c_{2} \p^{\m} h_{\l \r} \p^{\l} h^{\r \n}
+ c_{3} \p_{\l} h_{\r} ^{\ \m} \p^{\l} h^{\r \n} + c_{4} \p_{\l} h^{\l \m} \p_{\r} h^{\r \n} \nb \\
&+&  c_{5} \p_{\l} h^{\r \m} \p_{\r} h^{\l \n} + c_{6} \p^{\m} h^{\n \l} \p^{\r} h_{\l \r}
+ c_{7} \p^{\l} h^{\m \n} \p^{\r} h_{\l \r} + c_{8} \p^{\m} h^{\n \l} \p_{\l} h  \nb \\
&+& c_{9} \p^{\l} h^{\m \n} \p_{\l} h + c_{10} \p_{\l} h^{\l \m} \p^{\n} h
+ c_{11} \p^{\m} h \p^{\n} h + \eta^{\m \n} \big( c_{12} \p_{\t} h_{\r \l} \p^{\t} h^{\r \l} + \nb \\
&+& c_{13} \p_{\t} h_{\r \l} \p^{\r} h^{\t \l} + c_{14} \p_{\l} h^{\l \r} \p^{\t} h_{\t \r} + c_{15} \p_{\t} h^{\t \r} \p_{\r} h + c_{16} \p_{\r} h \p^{\r} h \big) + \nb \\
&+& b_{1} h^{\l \r} \p^{\m} \p^{\n} h_{\l \r} + b_{2} h \p^{\m} \p^{\n} h  + b_{3} h^{\m \n} \p_{\r }\p^{\r} h  + b_{4} h \p_{\r }\p^{\r} h^{\m \n} + \nb \\
&+& b_{5} h^{\m \l} \p_{\r} \p^{\r} h^{\n} _{\ \l}  +b_{6} h^{\r \n} \p^{\m} \p_{\r} h  + b_{7} h \p^{\m} \p_{\r} h^{\r \n}   + (\m \leftrightarrow \nu)  \ , 
\ea
where $h$ denotes the trace $\eta^{\m \n} h_{\m \n}$. Gauge invariance imposes nineteen conditions:
\ba
b_{1} = b_{2} = b_{3} = b_{4} = b_{5} = b_{6} = b_{7} &=& 0  \ , \\
c_{12} = c_{13} = c_{14}= c_{15} = c_{16} &=& 0  \ , \\
c_{7} = c_{8} = -c_{9} &=& - c_{6} \ , \\
c_{10} = - 2 c_{11} &=& - 2 c_{4} \ , \\
c_{2} &=& - 2 c_{1} \ , \\
c_{5} &=& c_{1} - c_{3} \ .
\ea
Which leaves $c_1,c_3,c_4$ and $c_6$ as possible non-zero independent coefficients. However, imposing $\p_{\m} \p_{\n} J_{h} ^{\m \n} = 0$ sets
\be
c_{1} = c_{3} = c_{4} = c_{6} = 0 \ .
\ee
As a result, there are no gauge-invariant self-couplings of the form (\ref{selfcouplingtotder}). We have also made an independent check by writing all possible two-derivative cubic terms in the action and imposing gauge invariance up to total derivatives, with the same outcome.

\end{document}